\documentclass[10pt,twocolumn]{aa}  

\usepackage{graphicx}
\usepackage{txfonts}
\usepackage{lipsum}
\usepackage{subcaption}     
\usepackage{linenoaa}    % necessary for continued figures, example in section 3
\usepackage{hyperref}

\begin{document}

   \title{Possible evidence for the 478 keV emission line from $^7$Be decay during the outburst phases of V1369 Cen}

   \titlerunning{V1369 Cen and $^7$Be 478 keV line}
   \authorrunning{Izzo et al.}

   \author{L. Izzo\inst{1,2}, T. Siegert\inst{3}, P. Jean\inst{4}, P. Molaro\inst{5,6}, P. Bonifacio\inst{7,5}, M. Della Valle\inst{1}, and T. Parsotan\inst{8}}

   \institute{INAF, Osservatorio Astronomico di Capodimonte, salita Moiariello 16, I-80131, Naples, Italy \\
             \email{luca.izzo@inaf.it}
            \and DARK, Niels Bohr Institute, University of Copenhagen, Jagtvej 155A, 2200 Copenhagen, Denmark
            \and Julius-Maximilians-Universit\"at Würzburg, Fakult\"at f\"ur Physik und Astronomie, Institut für Theoretische Physik und Astrophysik, Lehrstuhl f\"ur Astronomie, Emil-Fischer-Str. 31, D-97074 W\"urzburg, Germany
            \and Institut de Recherche en Astrophysique et Plan\'etologie, Universit\'e de Toulouse, CNRS, CNES, Universit\'e Toulouse III Paul Sabatier, 9 avenue Colonel Roche, BP 44346, 31028 Toulouse Cedex 4, France
            \and INAF, Osservatorio Astronomico di Trieste,, Via G.B.Tiepolo 11, Trieste, I-34143, Italy
            \and Institute of Fundamental Physics of the Universe, IFPU, Via Beirut, 2, Trieste, I-34151, Italy
            \and GEPI, Observatoire de Paris, Universite PSL, CNRS, 5 place Jules Janssen 92195 Meudon, France
            \and Astrophysics Science Division, NASA Goddard Space Flight Center, Greenbelt, MD 20771, USA}

   \date{Received YYY NN, 20XX}

% \abstract{}{}{}{}{}
% 5 {} token are mandatory
 
  \abstract
   {
   After decades of uncertainty about the origin of lithium, recent evidence suggests Galactic novae as its main astrophysical source. In this work, we present possible evidence for the first detection of the $^7$Be line at 478 keV, observed with the INTEGRAL satellite. The emission is temporally and spatially coincident with the outburst of the bright nova V1369 Cen, and line significance ranges from 2.5$\sigma$ to $\sim$1.9$\sigma$, depending on the detection methodology. A bootstrap analysis, assuming a fixed FWHM of 8 keV, provides a flux of $(4.9 \pm 2.0) \times 10^{-4}$ ph/cm$^2$/s centered at 479.0 $\pm$ 2.5 keV, with a 2.5$\sigma$ significant excess. This flux implies a total $^7$Be mass of $M_{^7Be} = (1.2^{+2.0}_{-0.6})$ $\times 10^{-8}$ M$_{\odot}$ at the distance determined using several indicators including the {\em Gaia} satellite. For a nova ejected mass estimated from radio observations, this result implies a $^7$Be=Li yield corresponding to $A(Li) = 7.1^{+0.7}_{-0.3}$. This value is comparable to those measured in a dozen novae through optical observations. Crucially, we confirm optically derived $^7$Li yields and demonstrate the groundbreaking potential of using gamma-ray data to measure Li abundances.
}


   \maketitle

%%%%%%%%%%%%%%%%%%%%%%%%%%%%%%%%%%%%%%%%%%%%%%%%%%%%%%%%%%%%%%
\section{Introduction}

The origin of lithium (Li) is one of the most puzzling problems in modern astrophysics. It is the heaviest element formed in the Big Bang nucleosynthesis (BBN) and the only primordial species for which there is a tension between observations and theoretical predictions \citep{1}. Moreover, the present Li abundance, as measured from meteorites or young T Tauri stars, is a factor of four larger than the BBN value and one order of magnitude larger than the abundance in old metal-poor stars \citep{2,3}. Galactic lithium sources are required to explain its present abundance \citep{4,5,6}. 
After decades in which astrophysicists have wandered in the dark with no concrete clues about the origin of Li, evidence has recently accumulated suggesting that Galactic novae might be the Li factories of the Universe.
One way to confirm that Li is produced in novae is by detecting the 478 keV emission line \citep{7} corresponding to the decay of beryllium-7 ($^7$Be) to Li through electron capture. Despite extensive searches, this long-sought-after emission line has still not been observed \citep{8}.

Classical novae (CNe) are recurring thermonuclear explosions that occur in binary-star systems consisting of a white dwarf (WD) that is accreting matter from a main sequence star or an evolved companion \citep{9,10}. When the pressure and temperature at the bottom of the accreted layer exceed the degeneracy pressure, thermonuclear reactions (TNR) ignite, removing degeneracy and causing the ejection of matter into the interstellar medium (ISM) \citep{11}. The main fuel of a CN explosion is represented by the CNO reaction cycle, which leads to a rapid increase in temperature and energy produced, and to the consequent ejection of a considerable amount of CNO isotopes into the ISM. During the TNR, the formation of beryllium-7 ($^7$Be) happens via the reaction $^3$He($\alpha$ , $\gamma$)$^7$Be, with $^7$Be being an unstable isotope that decays into $^7$Li via electron capture. The freshly-produced $^7$Be must be carried to the most external regions of the accreted layer by strong convective motions, which occur during the final stages leading up to the Nova explosion. This process, known as the \citet{12} mechanism, ensures that $^7$Be can survive the extreme conditions present in the hours or days immediately before the nova outburst.

$^7\rm{Be}$ decays by
electron capture through the reactions 
\begin{eqnarray}
{}^7\rm{Be}+ e^- \rightarrow {}^7\rm{Li}+ \nu_e \label{r1}\\
{}^7\rm{Be}+ e^- \rightarrow {}^7\rm{Li}^{\ast}+ \nu_e\label{r2}\\
{}^7\rm{Li}^{\ast}\rightarrow {}^7\rm{Li} + \gamma\label{r3}
\end{eqnarray}
where ${}^7\rm{Li}^{\ast}$ is a nuclearly excited state with
an energy of $477.6$ keV above the ground state. 
%  The Q-value of the reaction is $861.8$KeV \citep{tilley2002}.  
The laboratory value for the half-life time in reaction \ref{r2} %measured for neutral $^7$Be, 
is T$_{1/2} = 53.12 \pm 0.06$ days\footnote{Corresponding to a mean lifetime $\tau_{^7Be} = 76.64 \pm 0.09$ days} \citep{13}, while for reaction \ref{r3} the branching ratio is $10.52\%$\footnote{\url{http://www.escholarship.
org/uc/item/7p80t5p0}}. 
The detection of such a line associated with a nova outburst whose distance is known will provide a direct estimate of the Li yield \citep{14}. 

Several attempts have been made in the last decades to detect the 478 keV line from novae, using a variegated suite of high-energy detectors \citep{15,16}, which resulted only in upper limits on the flux of this line. More recently, \citet{8} used the INTEGRAL/SPI detector to search for evidence of the $\gamma$-ray lines originating from radioactive decays, including the 478 keV line, in the bright nova V5668~Sgr. In this nova, a large amount of $^7$Be II was found in near-UV spectra obtained within the first 90 days from the nova explosion \citep{17}. The lack of detection of the 478 keV line in INTEGRAL/SPI data is consistent with the distance of V5668 Sgr. One of the main outcomes of the work by \citet{8} is that any emission from the 478 keV line from nova explosions should be searched in events that are at closer distances, namely $d \leq 1kpc$.  

\section{The case of V1369 Cen}

V1369 Cen is currently the brightest classical nova explosion observed in this century: it was discovered on December 3rd, 2013, and it immediately reached the de-reddened magnitude V=3.3 mag \citep{18}. High-resolution optical spectra obtained seven days after its explosion have shown the presence of an absorption feature that was attributed to the resonance line of neutral lithium Li I 670.7 nm, at the same expanding velocity of $v_{\rm exp} = -550$ km/s, as the other line transitions identified for this nova \citep{18}. Here we revisit the peak luminosity and the distance of V1369 Cen, based in particular on the acquisition of more accurate, and multi-wavelength data, not available at the time of the nova explosion. 

\subsection{The maximum absolute magnitude of V1369 Cen}

\begin{figure*}[h!]
    \centering
    \includegraphics[width=0.32\linewidth]{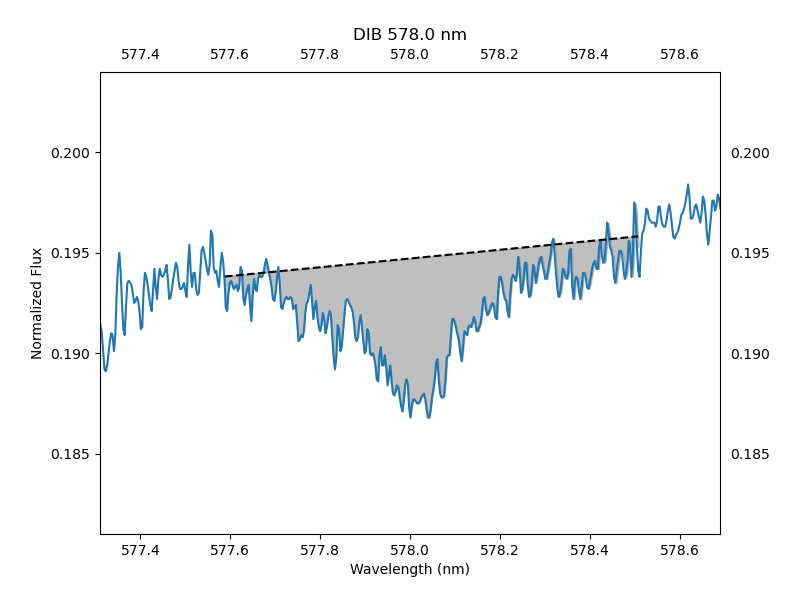}
    \includegraphics[width=0.32\linewidth]{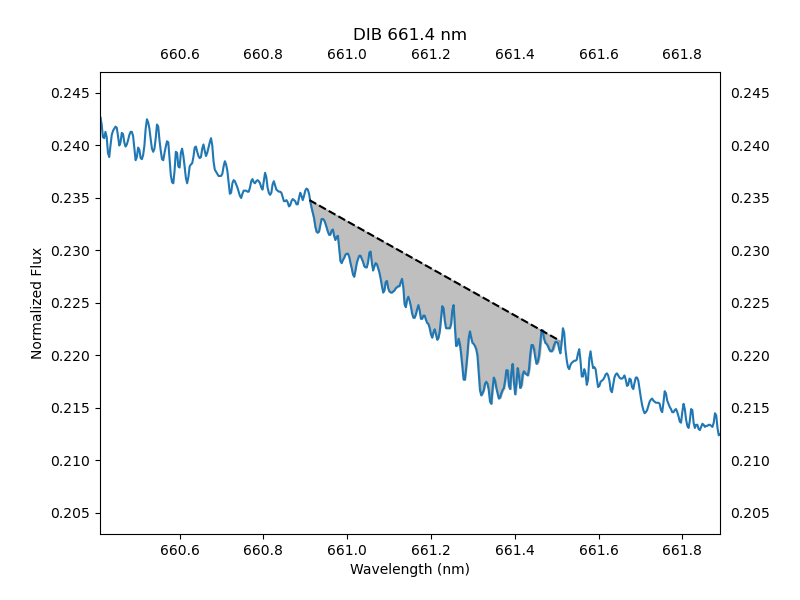}
    \includegraphics[width=0.32\linewidth]{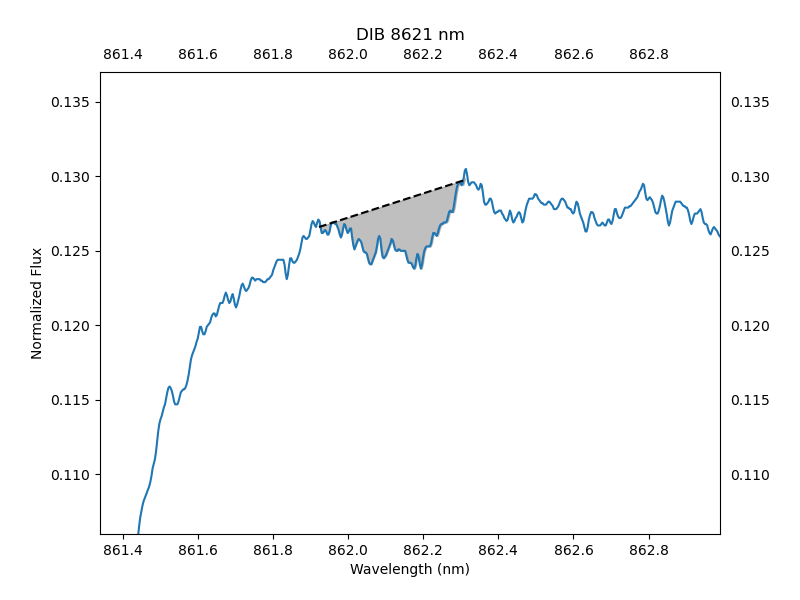}
    \caption{The DIBs identified in the early spectra of V1369 Cen and used for the estimate of the color excess $E(B-V)$. The gray area marks the region of the absorption line that has been used for the EW measurement, respectively for the DIB 578.0 nm (left panel), the DIB 661.4 nm (middle panel), and the DIB 862.1 nm (right panel).}\label{fig:DIB}
\end{figure*}

V1369 Cen has been extensively monitored by astronomers worldwide. The most detailed light curve of this nova was obtained using data collected by hundreds of amateur astronomers associated with the American Association of Variable Star Observers (AAVSO) \citep{19}. The V-band light curve from their observations shows a multi-peaked structure, with a peak brightness of V = 3.3 mag \citep{18}. This value is not corrected for the Galactic foreground extinction, $A_V$. The literature provides an $A_V$ value of 0.5 mag \citep{20}, based on the relation between the equivalent width of the interstellar Na ID lines and the color excess $E(B-V)$ \citep{21}. However, the Na ID IS lines in V1369 Cen spectra are saturated, indicating that the effective extinction is higher than this value (see also Appendix).  

An alternative method involves using diffuse interstellar bands (DIBs) detected in the early bright phases of V1369 Cen's spectra. DIBs are known to correlate well with the neutral hydrogen in the ISM, and their intensities serve as good tracers of the total line-of-sight color excess \citep{21,22,23,24}, similar to other ISM lines. To determine the overall extinction to V1369 Cen, we first identified the presence of the DIBs at 578.0 nm, 661.4 nm, and 862.0 nm in the early bright spectra of the nova. We then measured the equivalent widths (EWs) of the absorption features generated by each DIB using the FEROS spectra of V1369 Cen on Day 7 and Day 14. Fortunately, the 862.1 nm DIB is located in a spectral region free from strong telluric lines. Fig. \ref{fig:DIB} shows the identification and the region used for the EW measurement for each DIB, and Table \ref{tab:DIB} reports our measurements.

The color excess $E(B-V)$ is determined from DIB EWs using empirical correlations published in the literature, derived from high-resolution spectra of large samples of Galactic stars of all spectral types. The DIB at 578.0 nm is widely used in the literature due to its strong presence in stellar spectra. We refer to correlations found in large samples of early-type local ($\sim$300 pc) stars with high-quality spectra \citep{22}, in low-resolution (R $\sim$ 3,000) SDSS and LAMOST spectra of Galactic stars exhibiting a wide range of extinction \citep{26}, and in young stellar objects \citep{23}. This latter sample was also used to correlate the DIB at 661.4 nm with color excess, supported by a detailed study from the {\em Gaia}-ESO collaboration linking the EW of this DIB with total extinction along the line of sight in cool star spectra \citep{27}. The DIB at 862.1 nm is one of the best tracers of the Galactic ISM spatial structure \citep{25} and interstellar reddening, showing a tight correlation with the color excess along the line of sight of several stars \citep{22,26} and a clear correspondence with Galactic CO gas velocities \citep{23}.

Using these extinction correlations and the DIB EW measurements from Table \ref{tab:DIB}, we calculated a list of color excesses. The weighted average of these values gives $E(B-V) = 0.29 \pm 0.04$ mag. Assuming a \citet{28} extinction curve, this results in a total V-band extinction of $A_V = 0.90 \pm 0.12$ mag. Consequently, the de-reddened peak brightness of V1369 Cen is $V_{max} = 2.4 \pm 0.1$ mag.

 \begin{table*}[h!]
    \centering
    \begin{tabular}{l|c|c|c|c}
    \hline\hline
       DIB  & EW & Epoch & $E(B-V)$ & Ref.\\
       (nm) & (m\AA) & (Days) & (mag) &  \\
        \hline
      578.0 & 140 $\pm$ 7 & 13  & 0.27 $\pm$ 0.02 & \citep{26}\\
      578.0 & 140 $\pm$ 7 & 13  & 0.31 $\pm$ 0.07 & \citep{22}\\
      578.0 & 140 $\pm$ 7 & 13  & 0.26 $\pm$ 0.02 & \citep{21}\\
      661.4 & 72 $\pm$ 12 & 13  & 0.70 $\pm$ 0.10 & \citep{26}\\
      661.4 & 72 $\pm$ 12 & 13  & 0.35 $\pm$ 0.09 & \citep{22}\\
      862.1 & 75 $\pm$ 4 & 7  & 0.29 $\pm$ 0.06 & \citep{27}\\
      \hline\hline
    \end{tabular}
    \caption{The measurement of DIB's EWs (columns 1 and 2) in the early optical spectra of V1369 Cen (column 3), and the corresponding E(B-V) value (column 4) obtained using the methodology cited in column 5.}
    \label{tab:DIB}
\end{table*}

\subsection{The distance to V1369 Cen}

The distance to V1369 Cen is the most important parameter of the nova for which we do not have a precise estimate. Recent analyses suggest a range between $d = 1.0 \pm 0.4$ kpc, based on updated 3D Galactic reddening maps \citep{29}, and up to 2.5 kpc, inferred from ISM lines and the H I 21 cm line profile \citep{30}. V1369 Cen was observed by the {\em Gaia} satellite \citep{31} multiple times, with data release 3 (DR3) covering observations between July 25, 2014, and May 28, 2017 \citep{32}. During this period, V1369 Cen was observed in 60 visits. The resulting parallax is $p = 3.74 \pm 0.74$ mas, and the distance inferred by a detailed Bayesian treatment of {\em Gaia} DR3 data is $d =  643 (+405, -112)$ pc \citep{33}. 

Interestingly, the distance derived using the General Stellar Parametrized from Photometry (GSP-Phot) methodology yields a much larger value. However, this method relies on {\em Gaia} Bp/Rp spectra matched to synthetic spectra from astrophysical models. The {\em Gaia} spectra of V1369 Cen, dominated by nebular spectral features from the 2013 outburst, do not resemble stellar templates, indicating that the GSP method is not applicable for V1369 Cen. Additionally, given the extinction-corrected peak magnitude $V_{max} = 2.4 $ mag, the derived absolute magnitude at the GSP distance would be $M_V = -10.7$ mag, much brighter than typical for very fast novae \citep{10}. V1369 Cen, a moderately slow nova with $t_2 = 40 \pm 5$ days \citep{18}, suggests a fainter absolute magnitude.

\begin{figure}
    \centering
    \includegraphics[width=0.88\linewidth]{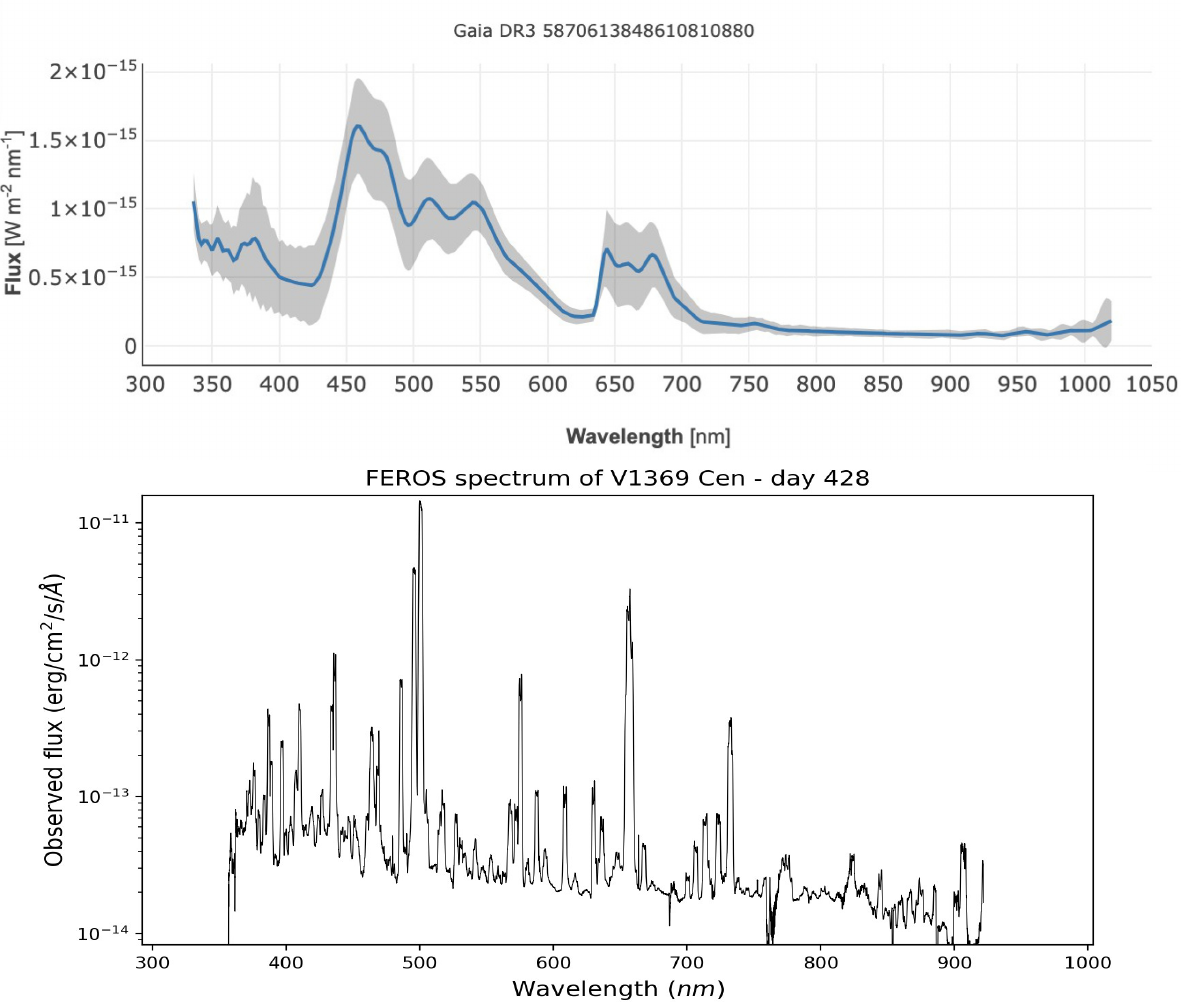}
    \caption{(Upper panel) The low-resolution {\em Gaia} spectrum of V1369 Cen obtained in 2015. (Lower panel) The high-resolution FEROS spectrum of V1369 Cen obtained on Feb 3rd, 2015. The spectral features observed in the {\em Gaia} spectrum match very well the emission lines of the FEROS spectrum, indicating that the nova was in a similar spectral phase, the nebular stage, and then that the spectrum does not have stellar features. This implies that the GSP methodology for the derivation of the distance of V1369 Cen could not work for the case of V1369 Cen.}
    \label{fig:S2}
\end{figure}

At the DR3 distance, the absolute magnitude at maximum would be $V_{max} = -6.6 (+1.1, -0.3)$ mag, which is in agreement with expectations from the Maximum Magnitude and Rate of Decay (MMRD) relation \citep{10}. This relation links the absolute peak brightness of a nova with its decay rate, parameterized by the time a nova decays by two (to three) magnitudes, namely $t_2$ ($t_3$). Assuming the measured de-reddened peak brightness $V_{peak} = 2.4 \pm 0.1$ mag, and $t_2 = 40 \pm 5$ days, we find that to conform to the MMRD relation within 2$\sigma$, V1369 Cen must be within 550 pc to 1400 pc. Notably, the largest distance reported in the literature \citep{30} is more than 3$\sigma$ off the MMRD relation.

\begin{figure}
    \centering
    \includegraphics[width=0.98\linewidth]{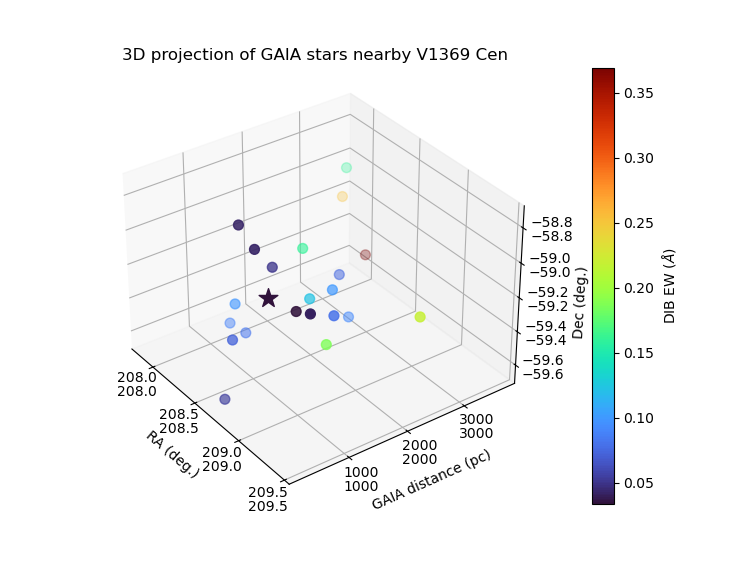}
    \caption{The 3D distribution of the sample of 21 {\em Gaia} DR3 stars surrounding V1369 Cen. The third spatial dimension is provided by the {\em Gaia} distance, while the color of different markers is related to their corresponding DIB EW values. V1369 Cen is represented with a star marker, while the other GAIA DR3 stars are reported with circle markers. }
    \label{fig:3D}
\end{figure}

We have also used an alternative approach, which is based on the correlation between the Diffuse Interstellar Band (DIB) at 862.1 nm and color excess for nearby stars. Distant stars exhibit more interstellar reddening, resulting in larger DIB EW values. We built a correlation between DIB EW and {\em Gaia} DR3 distance for stars near V1369 Cen. The {\em Gaia} collaboration has employed a similar method to study the Galactic ISM using the DIB at 862.1 nm in the RVS passband \citep{34}. However, RVS spectra from {\em Gaia} DR3 are available only for stars brighter than 14 mag. From an initial sample of 625 stars, we identified 45 with {\em Gaia}-RVS spectra, of which only 21 were reliable for analysis due to a pipeline issue affecting 24 stars. The distribution of these stars, along with V1369 Cen at 970 pc, is shown in Fig. \ref{fig:3D}, with marker color indicating DIB 862.1 nm values.

\begin{figure}
    \centering
    \includegraphics[width=0.98\linewidth]{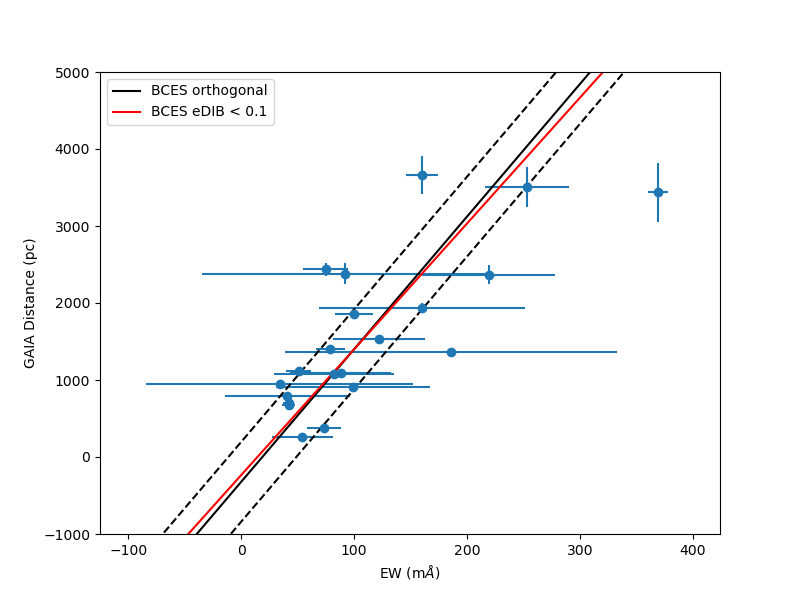}
    \caption{The DIB 862.1 nm EW vs {\em Gaia} distance distribution. The black curve represents the best fit found from the entire sample of 21 stars, while the red curve refers to the same analysis performed on stars characterized by an uncertainty on the EW $<$ 100 m$\AA$.}
    \label{fig:Gaia_comp}
\end{figure}

The distribution of DIB EW vs {\em Gaia} DR3 distance for these 21 stars is shown in Fig. \ref{fig:Gaia_comp}. We performed a best-fit analysis considering uncertainties on both DIB EW and {\em Gaia} distance, including an intrinsic scatter parameter. Using the 'orthogonal' method from the {\sc BCES} python package \citep{43}, we found a correlation between DIB 862.1 nm EW and {\em Gaia} distance, shown as a black curve in Fig. \ref{fig:Gaia_comp}, with 2$\sigma$ uncertainty in dashed lines. Excluding three stars with DIB EW uncertainties larger than 0.1 m$\AA$ did not significantly affect the result, shown as a red curve in the same figure. The derived distance for V1369 Cen from the DIB measurement is $d_{V1369Cen} = 970.4 \pm 460.3$ pc, consistent with \citet{29} and the {\em Gaia} DR3 distance \citet{32}. Based on all the considerations reported above, we consider the distance to V1369 Cen the value reported in \citet{33}.

\begin{figure}
    \centering
    \includegraphics[width=0.88\linewidth]{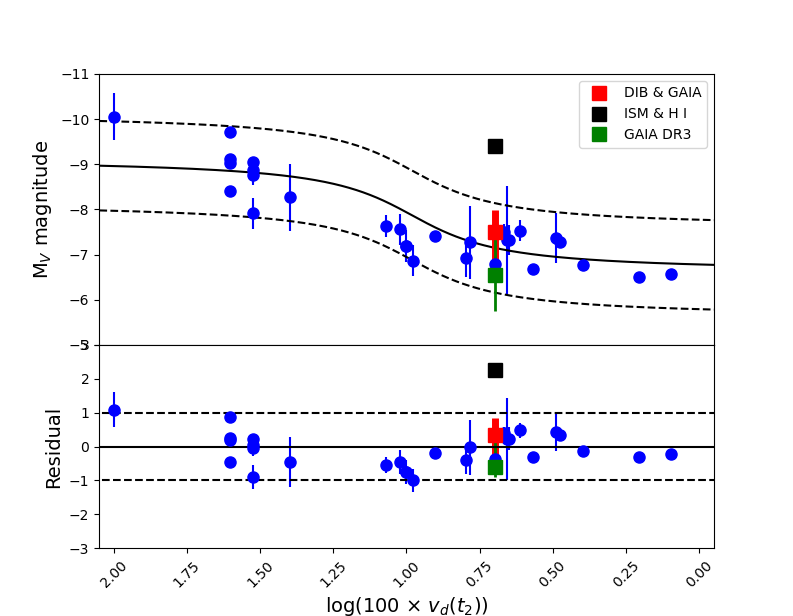}
    \caption{The Galactic MMRD relation estimated using the sample of novae whose distance has been measured with {\em Gaia} \citep{10}. The dashed lines correspond to 2$\sigma$ confidence region. The red data marks the position of V1369 Cen using the distance derived from the use of the DIB 862.1 nm and {\em Gaia} stars in the surroundings of V1369 Cen. tThe black data marks the position of V1369 Cen for a distance of 2.4 kpc \citep{30}, while the green data represents the position of V1369 Cen assuming a {\em Gaia}-DR3 distance \citep{33}.}
    \label{fig:MMRD}
\end{figure}

\section{INTEGRAL observations of V1369 Cen}

INTEGRAL has been observing the gamma-ray sky since its launch in 2002 \citep{35}. During this time, more than a hundred Galactic novae have been detected in outbursts at optical wavelengths\footnote{\url{https://asd.gsfc.nasa.gov/Koji.Mukai/novae/novae.html}}. This number reduces to half its original value since {\em Gaia}, which is a satellite dedicated to measure parallaxes and proper motions of billions of stars in the Milky Way including novae, has been operational \citep{31}. We searched in the INTEGRAL archive for observations of the region in the sky where V1369 Cen was located, using a search radius value of $\sim$ 10 degrees. 

INTEGRAL has observed in multiple visits the region of the sky surrounding the location of V1369 Cen. In particular, a dedicated target of opportunity observations was performed 24 days after the nova discovery in order to follow up the gamma-ray detection of the nova by the Large Area Telescope detector on board the Fermi spacecraft \citep{36}. The list of INTEGRAL revolutions for which V1369 Cen is within the partially coded field of view of the Spectromètre Pour Integral (SPI), namely the angular distance to the pointing axis $\lesssim 16$ deg, is reported in Table \ref{tab:S1}. This table in particular reports the time duration within the angular distance to the nova versus the detector direction to the sky, which is lower than 31 degrees. V1369 Cen was never in the fully coded field of view for revolutions with $R_{min} > 8$ deg, and this implies that the effective area of the detector is reduced for those observations. In particular, during the revolution 1368, on December 27, namely 24 days after the nova discovery, which was a dedicated ToO activation to observe gamma-ray transients (prog. ID 1040030, PI: den Hartog), the angular distance of the direction to V1369 Cen and the pointing axis was indeed very small, slightly variable during the entire duration of the observations ($t_{exp} = 90$ ks) from 2.7 to almost zero degrees. Observations were executed using a hexagonal pattern.

\begin{table}[]
    \centering
    \begin{tabular}{l|c|c|c|c|c}
    \hline\hline
       Revolution  & Nscw & $\Delta t$ & $T_{elapsed}$ & $R_{min}$ & $R_{max}$\\
        & & (Days) & (s) & (degrees) & (degrees) \\
        \hline
      1364 & 2 & 14.0 & 18529 & 10.3 & 10.5\\
      1365 & 9 & 14.9 & 20760 & 10.2 & 14.2\\
      1366 & 10 & 18.0 & 10812 & 5.1 & 11.6\\
      1368 & 53 & 24.0 & 149807 & 0.0 & 2.7\\
      1370 & 11 & 30.7 & 35328 & 11.7 & 14.8\\
      1371 & 12 & 33.7 & 35362 & 11.7 & 15.0\\
      1378 & 28 & 53.9 & 70428 & 10.4 & 15.4\\
      1379 & 22 & 57.2 & 50884 & 10.5 & 15.4\\
      1386 & 6 & 78.7 & 21206 & 11.1 & 15.0\\
      \hline\hline
    \end{tabular}
    \caption{The diary of INTEGRAL/SPI observations showing revolutions when V1369 Cen was in the field of view of the detector. The columns report the revolution number, the number of science windows, the days after the discovery date $\Delta t$,  the elapsed time for which the angular distance is lower than 16 degrees ($T_{elapsed}$), and the minimum and maximum angular distance to the nova ($R_{min}$, and $R_{max}$), respectively.}
    \label{tab:S1}
\end{table}

\begin{figure}
    \centering
    \includegraphics[width=0.98\linewidth]{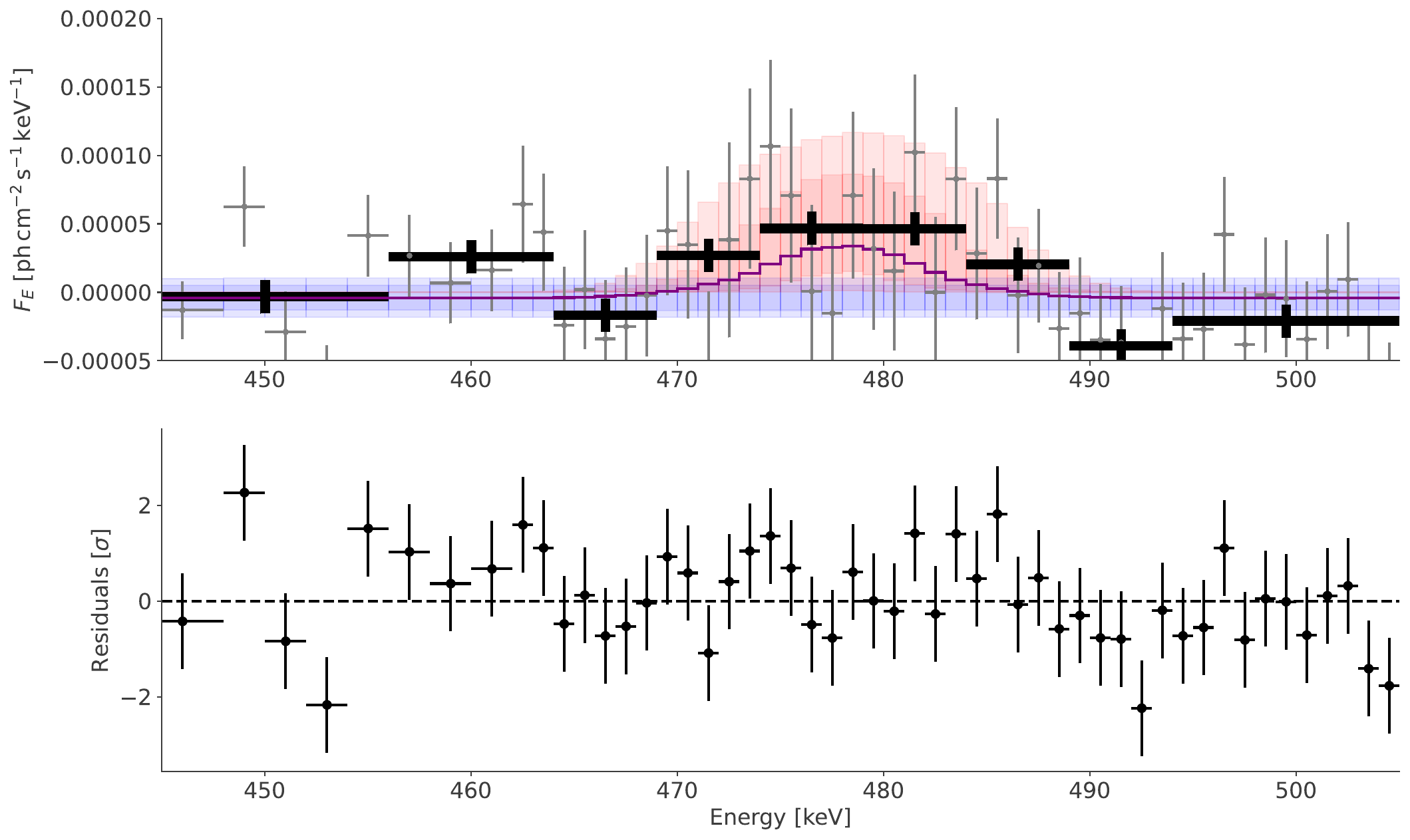}

    \caption{The detection of the 478 keV line in INTEGRAL/SPI data during revolution 1368, obtained with the analysis using a fixed Gaussian line width. Inspection of the full spectrum from 20 keV to 505 keV shows that it is consistent with zero everywhere except for the bump at 478 keV. The top panel shows the extracted fluxes (gray) and rebinned (black). The fitted spectrum is shown as a blue (constant) and red (line) band with their 1 and 2 sigma uncertainties. The bottom panel shows the residuals of the fit for the top panel (the plot for the line with a free line width is very similar).}
    \label{fig:det}
\end{figure}

\begin{figure}
    \centering
    \includegraphics[width=0.98\linewidth]{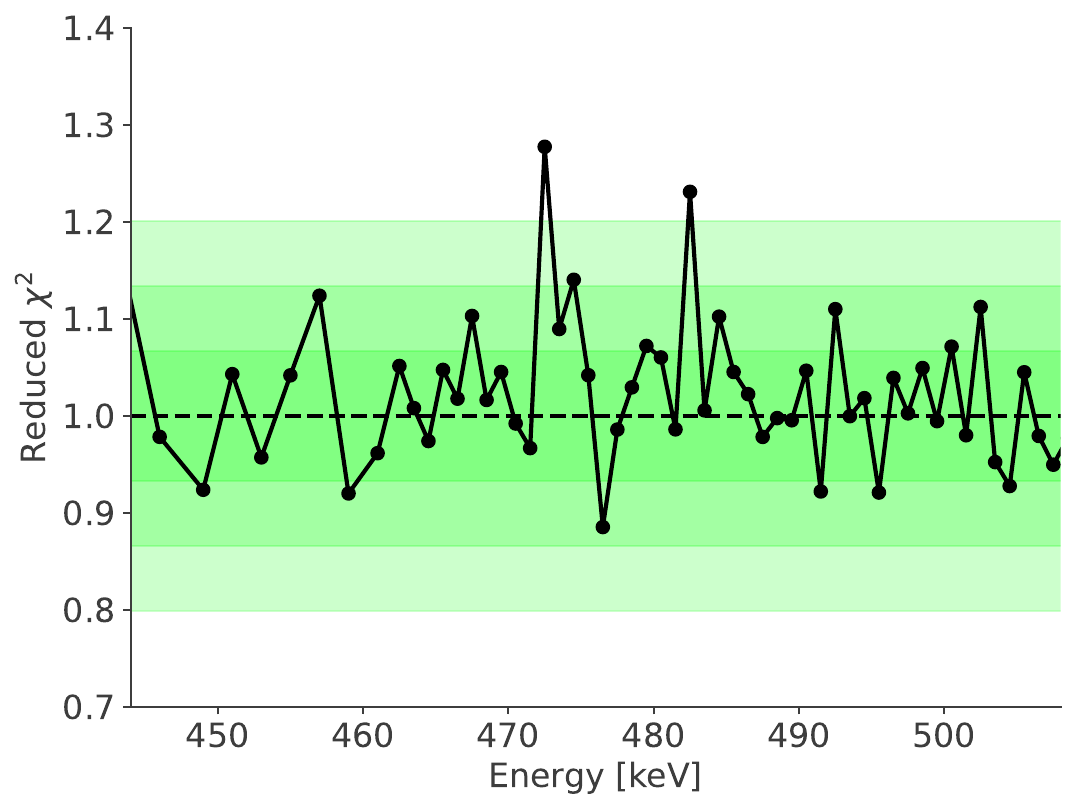}
    \caption{The reduced $\chi^2$ values measured within the range surrounding the 478 keV line, for the fixed Gaussian model case.}
    \label{fig:S0}
\end{figure}

\begin{figure}
    \centering
    \includegraphics[width=0.98\linewidth]{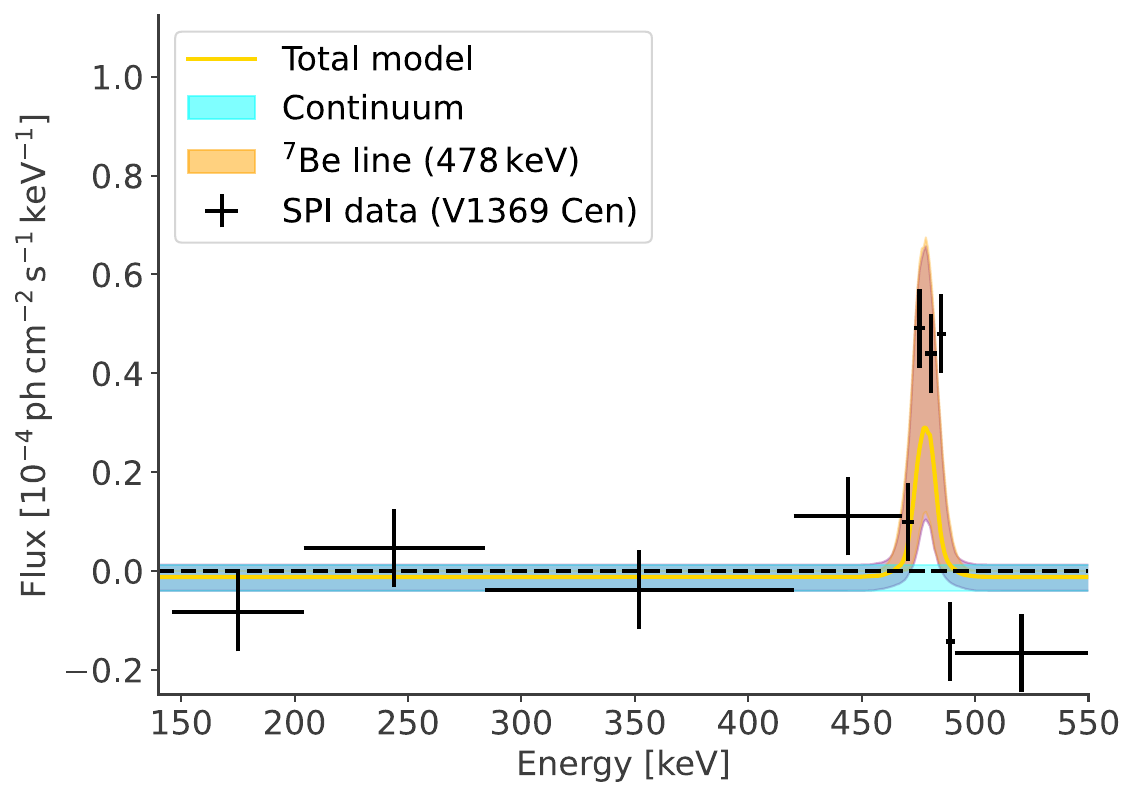}
    \caption{Broadband spectrum of the position of V1369 Cen in INTEGRAL revolution 1368. Shown are extracted spectral data points from 150 keV to 550 keV which are all consistent with zero except around the 478 keV line.}
    \label{fig:S1}
\end{figure}

We have performed a detailed analysis of INTEGRAL/SPI data of V1369 Cen for INTEGRAL revolution 1368. V1369 Cen was observed for about 90 ks in a hexagonal pattern with 15 among 19 active detectors. The gamma-ray spectrum was extracted from SPI/INTEGRAL data by a model-fitting method, which consists in fitting the flux of the source and the instrumental background rate, in each energy bin, to the counts measured per pointing and per detector. The instrumental background rate was fitted using two different methods \citep[][but see also section 1.3 of \citealp{8} and the method {\sc ORBIT-DETE} in section 3 of \citealp{56}]{55} yielding a difference in the flux of $\approx$ 11$\%$ ($\approx$ 0.28 $\sigma$). This systematic difference is lower than the other statistical and systematic uncertainties (e.g. distance, date of the thermonuclear runaway) and will not be taken into account in the following (see also below). 
We started an analysis of the data from 20 keV to 505 keV, and found that the spectrum is consistent with zero everywhere, except for a 2.5 $\sigma$ bump exactly at 478 keV, see Fig.\ref{fig:det}. The reduced $\chi^2$ values in the range of the 478 keV line are displayed in Fig. \ref{fig:S0}. The flux in the remaining INTEGRAL/SPI range between 20 to 400 keV is consistent with zero flux, see also Fig. \ref{fig:S1}. The significance of this detection is strongly affected by the relatively short exposure time used during revolution 1368, which was the only observation where the nova's location was fully centered within the coded field of view of the SPI detector. To further assess systematics, we employed a third method that involved fitting a scaling factor to a fixed detector pattern \citep{38} for each pointing and energy bin. This approach yielded a slightly lower significance with a difference of -0.59 $\sigma$ compared to the chosen value. In this method, the detector pattern was obtained using the relative background count rate between detectors measured per orbit for each energy bin. Based on the above analyses, we conclude that the line significance varies from $\sim$ 1.9 $\sigma$ to 2.5 $\sigma$, depending on the chosen background determination method.

Then we fit the spectrum in the restricted range between 445 and 505 keV, which includes the 478 keV line, using a constant model and an additional Gaussian line.
We have employed two different analyses. In the first analysis, we fixed the FWHM of the Gaussian line to the value of 8 keV FWHM, according to \citet{8}, obtaining an integrated photon flux for the line of $F_F = (4.9 \pm 2.0) \times 10^{-4}$ ph/cm$^2$/s, with line center at 479.0 +/- 2.5 keV. 

\begin{figure}
    \centering
    \includegraphics[width=0.98\linewidth]{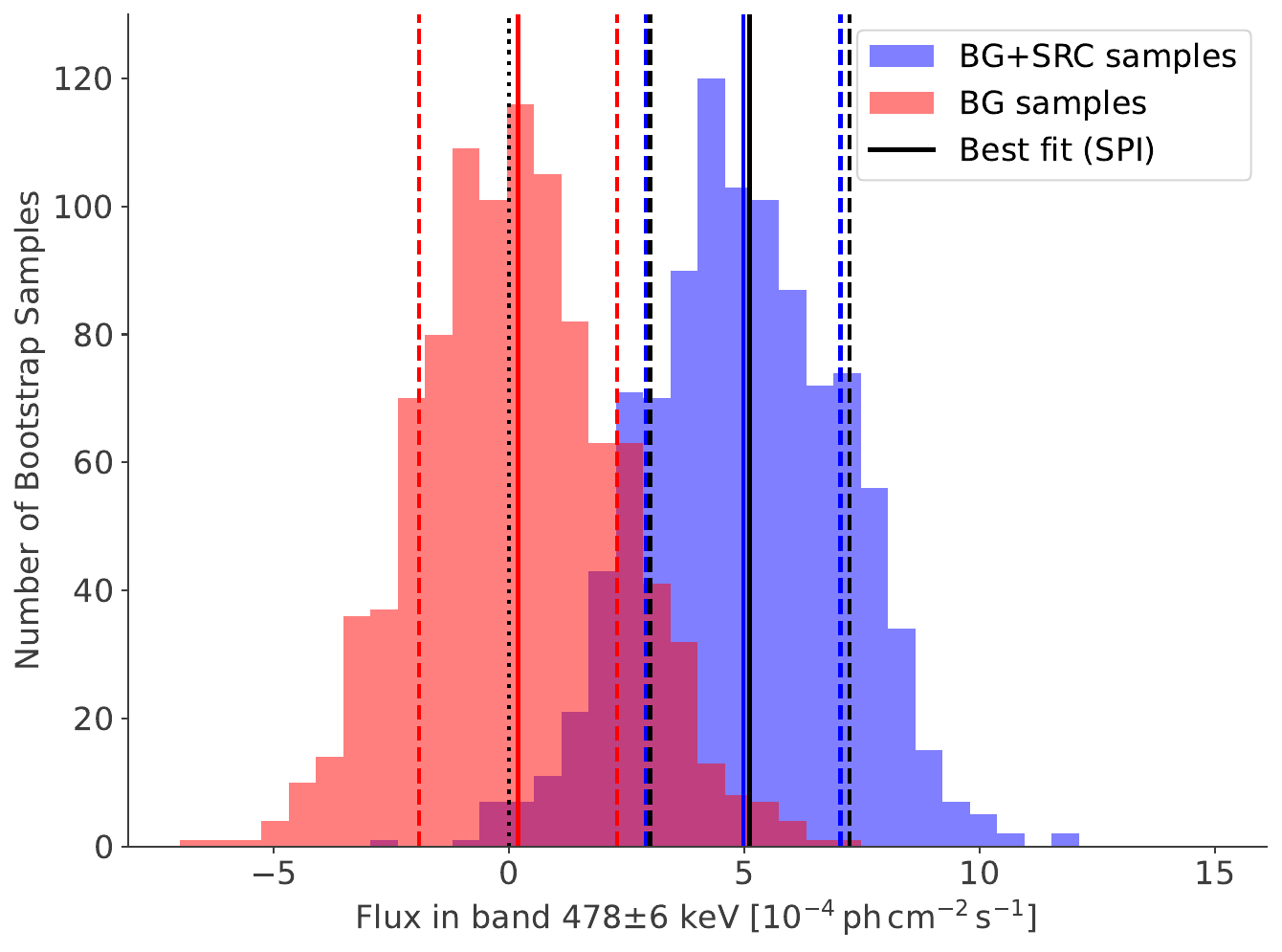}

    \caption{ The results of the bootstrap analysis described in the text. The red histogram represents 1000 bootstrap resampled datasets from the INTEGRAL/SPI containing background data, while the blue histogram illustrates samples from a band centered at 478.0 keV with a 12 keV width. Comparing these two distributions reveals a $\approx$ 2.5$\sigma$ significance level for the observed sky flux exceeding the background, indicating a substantial detection of the spectral line.}
    \label{fig:boots}
\end{figure}

In order to more accurately evaluate the significance of the flux excess, we generated 1000 bootstrap samples using data from the SPI detector and the background flux in a 12 keV-wide band centered at 478.0 keV. This bandwidth corresponds to the full width at half maximum (FWHM) of approximately 8 keV for a Gaussian line. By comparing the two resulting distributions, we confirmed a 2.5$\sigma$ significance level for the observed flux excess, see Fig. \ref{fig:boots}.

\begin{figure}
    \centering
    \includegraphics[width=0.98\linewidth]{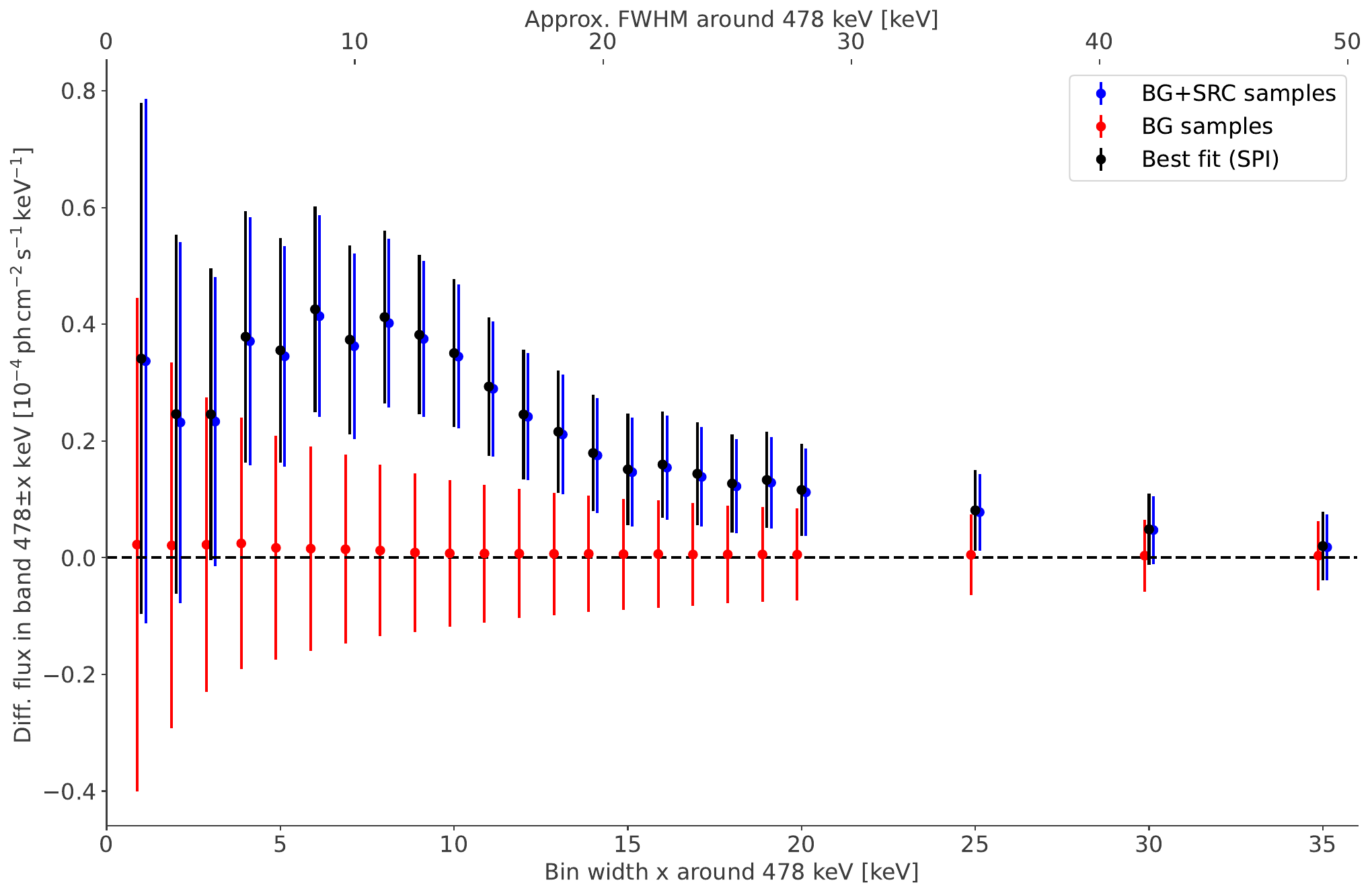}

    \caption{ The distribution of differential line fluxes normalized to the different extraction band width around 478.0 keV. Red data gives the result for the background region alone, while blue data corresponds to an extraction bin centered at 478.0 keV. The variation of the flux for different bin widths, with a maximum reached at about 8 keV suggests that only for broad line widths we get a significant signal for the line.  }
    \label{fig:diffflux}
\end{figure}

In the second analysis, we have relaxed the constraint on the width of the line, obtaining an integrated photon flux of $F_T = (6.9 \pm 3.0) \times 10^{-4}$ ph/cm$^2$/, and a width of $FWHM = 3200$ km/s. Figure \ref{fig:diffflux} displays the normalized flux as a function of varying slit width. The results illustrate that a slit with a FWHM of approximately 8 keV yields the maximum flux. Consequently, the observation of the spectral line is exclusively detectable at this specific slit size.

Using the formula Eq. \ref{eq:2}, we can convert this estimate in the initial mass of $^7$Be that was synthesized in the outburst, after considering a delay time of $t = 24$ days from the nova discovery. Assuming a {\em Gaia} distance for V1369 Cen, we obtain a total $^7$Be mass of $M_F = 1.2^{+2.0}_{-0.6}$ $\times 10^{-8}$ M$_{\odot}$, while in the case of a relaxed constrain on the FWHM of the Gaussian line, we measure a total synthesized mass of $M_T = 1.6^{+2.8}_{-0.9}$ $\times 10^{-8}$ M$_{\odot}$.

\begin{figure}
    \centering
    \includegraphics[width=0.98\linewidth]{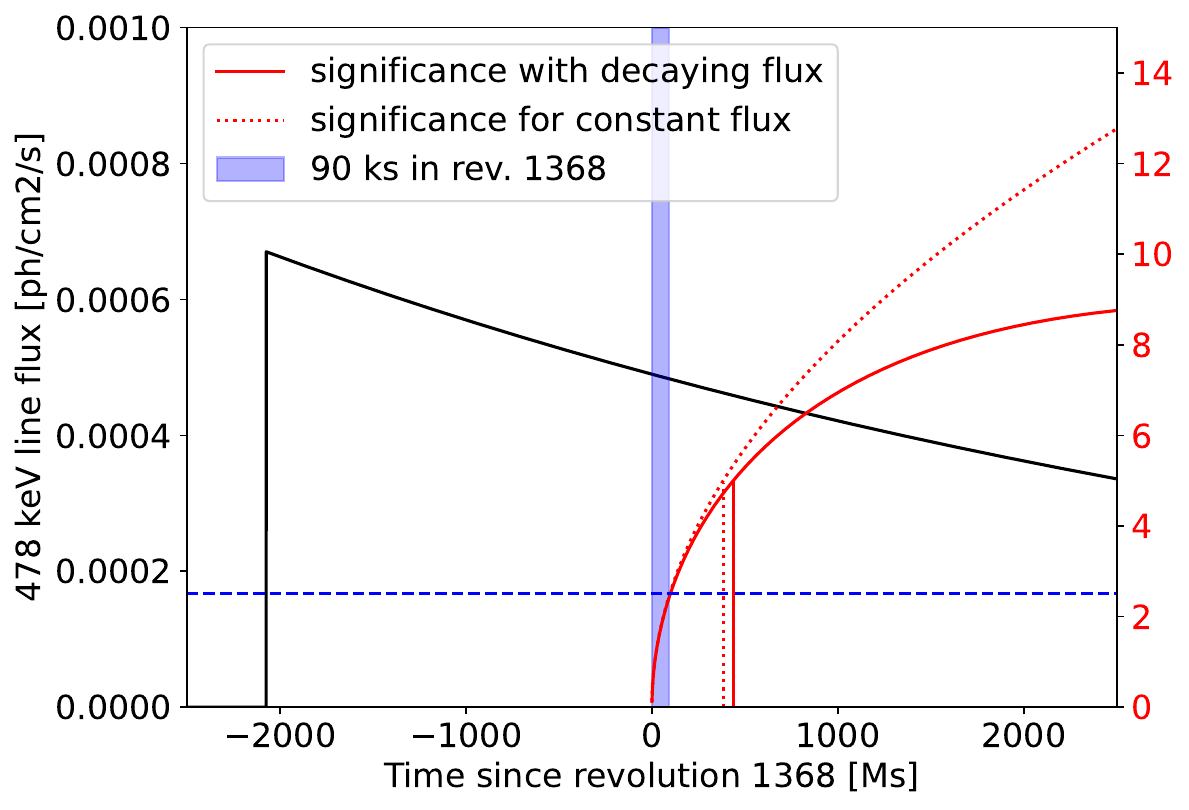}
    \caption{The effective exposure time required to achieve a 5$\sigma$ detection of the 478 keV line was determined based on the observed flux during revolution 1368 and the estimated explosion time of the nova. Assuming a constant flux over time for the 478 keV feature (red curve) and a decaying flux following the time decay of $^7$Be (red dashed curve), we find that an uninterrupted exposure time of 440 ks would be necessary. However, due to instrumental constraints (see text), this corresponds to an effective exposure time of approximately 780 ks to reach a 5$\sigma$ detection. }
    \label{fig:S1b}
\end{figure}

To determine the effective exposure time needed to achieve a 5$\sigma$ significance detection for the 478 keV emission line, we performed simulations based on the flux measured during revolution 1368 and the expected explosion time of the nova. We considered two scenarios: (1) a constant line flux over time and (2) a line flux that decays over time according to the mean lifetime $\tau_{^7Be}$ of the $^7Be$ isotope. Our results indicate that an exposure time of $t = 440$ ks would have been required to achieve a 5$\sigma$ significance detection for the 478 keV line (see Fig. \ref{fig:S1b}). However, since the SPI detector can observe only approximately 85$\%$ of the 3-day INTEGRAL orbit, the effective exposure time needed to reach this significance level would be $t_{\text{eff}} \sim 780$ ks.

\subsection{Swift Burst Alert Telescope Observations and Upper Limits}

We used the BatAnalysis python package \citep{39} to analyze Swift Burst Alert Telescope (BAT) data from 2013-11-03 to 2013-12-10. We analyzed survey data, where the location of V1369 Cen had at least a partial coding fraction of $\sim 19\%$ on the BAT detector plane. The total set of observations that BAT took in the time period from 2013-11-03 to 2013-12-10 amounted to $\sim 2,581,303$ s of exposure time, while the coordinates of V1369 Cen had a $\sim 19\%$ partial coding or greater for $149,743$ s of exposure time. Thus, BAT was observing the target with a partial coding of $\gtrsim 19\%$ for $\sim 7\%$ of the time.

We additionally constructed daily mosaiced images using the package to obtain potentially more significant detections of the nova. Overall, there was no significant detection of the nova in the BAT survey or daily mosaiced data. Using the BatAnalysis tool, we are also able to place upper limits on the nova emission in the 14-195 keV energy range for each survey and mosaic dataset. We find that the flux upper limit in the 14-195 keV energy range is $\lesssim 6 \times 10^{-9}$ erg/cm$^2$/s 12 days before the nova was detected. 

\begin{figure}
    \centering
    \includegraphics[width=\linewidth]{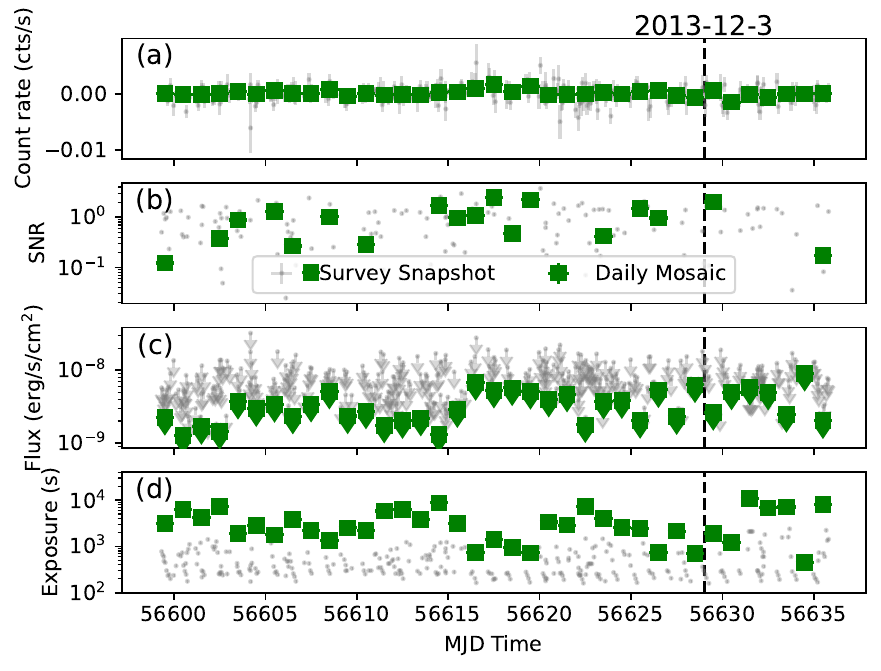}

    \caption{Swift BAT observations of the location of V1369 Cen from 2013-11-03 to 2013-12-10. No significant detections of the nova was made but upper limits can be placed on the 14-195 keV emission at the level of $\lesssim 6 \times 10^{-9}$ erg/cm$^2$/s, 12 days before the nova was detected. }
    \label{fig:BAT}
\end{figure}

In Figure \ref{fig:BAT} we show the count rate of the nova in panel (a), the measured SNR where it can be reliably determined in panel (b), the flux upper limits in panel (c), and the total exposure time of the source in panel (d). The gray points denote the survey data-derived quantities and the green points show the quantities obtained from the daily mosaics. 

\section{Discussions}

The mass of $^7$Be synthesized in the TNR can be directly derived from the observed INTEGRAL/SPI photon flux $F$, using the following formula \citep{8}:

\begin{equation}\label{eq:2}
M_{^7Be} = F_{^7Be} \frac{4 \pi d^2 m_{^7Be} \tau_{^7Be}}{ p_{^7Be}} \exp{ \frac{t + \Delta t}{\tau_{^7Be}}},
\end{equation}

where $m_{^7Be} = N_{^7Be} \times u$ is the $^7$Be atomic mass, and $u$ is the atomic mass unit value. The quantity $\Delta t$ represents the delay time between the ignition of the TNR and the moment when the ejecta becomes optically thin to $\gamma$-ray photons \citep{8} which is not well known, but likely to be on the order of a few days. We set this parameter to 5 days. Using Eq. \ref{eq:2}, at the adopted distance of V1369 Cen, and considering $t - \Delta t$ = 24 $\pm$ 5 days, we obtain a total $^7$Be mass of $M_{^7Be} = (1.2^{+2.0}_{-0.6})$ $\times 10^{-8}$ M$_{\odot}$. In the case of the thawed FWHM, we measure a total synthesized mass of $M_{^7Be} = (1.6^{+2.8}_{-0.9})$ $\times 10^{-8}$ M$_{\odot}$. 

\begin{figure}
    \centering
    \includegraphics[width=0.98\linewidth]{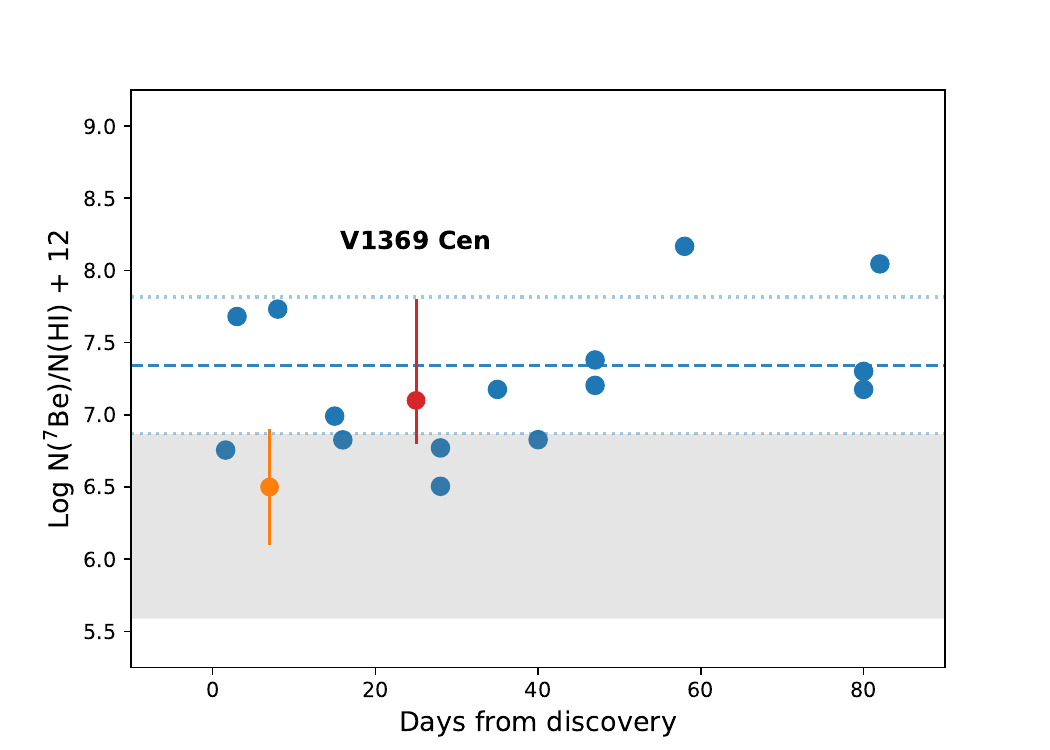}
    \caption{The $^7$Be yields measured in classical and recurrent novae from analysis of the near-UV $^7$Be II 313.0 nm line (blue data). The plot shows the atomic ratio of $^7$Be to H:  N($^7$Be) = $\log{N(^7Be)/N(H I)}$ + 12,  on the y-axis. For V1369 Cen, the yield measured from the 478 keV line is N($^7$Be) = 7.1$^{+0.7}_{-0.3}$ (red data), whereas the value, adjusted for the new GAIA distance, from the optical data \citep{33} is N($^7$Be) = 6.5 $\pm$ 0.4 (orange data). This latter value is lower by 0.6 dex compared to the 478 keV analysis but still falls within the margin of error. The dashed line indicates the average value ($\pm$ 1$\sigma$), derived from the entire sample of novae. The gray area represents the theoretical predictions from various models \citep{46,47,48}.}
    \label{fig:2}
\end{figure}

The ejected mass in V1369 Cen has been obtained using data obtained with the Australian Square Kilometer Array Pathfinder (ASKAP) during a systematic survey performed to search radio counterparts of classical novae \citep{40}. The ejected mass of V1369 Cen, at the distance of $d = 1.0 \pm 0.4 kpc$ \citep{29}, which is similar to the distance adopted in this work, is $M = (1.65 \pm 0.17) \times 10^{-4}$ M$_{\odot}$. However, this mass value has been derived using a pure hydrogen composition for the nova ejecta in V1369 Cen. A more realistic assumption consists in considering a contribution from helium to the electron density population responsible for the observed radio emission. In the Hubble flow model for nova shells emitting at radio frequencies \citep{41} a plasma with singly ionized helium and hydrogen, with a numerical abundance ratio of 0.15 is assumed.  The contribution from heavier particles is negligible, given that the abundance of these elements in nova ejecta is of the order of 10$^{-3}$ - $10^{-4}$ times lower than hydrogen \citep{42}. Considering this abundance ratio, and their density derived from ASKAP radio data, we have determined the hydrogen and helium masses ejected in V1369 Cen to be $M_{ej,H} = (1.40 \pm 0.14) \times 10^{-4}$ M$_{\odot}$ and $M_{ej,He} = (2.47 \pm 0.25) \times 10^{-5}$ M$_{\odot}$. With these values, and the $^7$Be mass found from analysis of the 478 keV line, we obtain a total lithium yield of $\log N(^7Be)/N(H) + 12 = 7.1^{+0.7}_{-0.3}$, a value that is fully consistent with the average novae Li yield of $\log N(Li)/N(H) + 12 = 7.34 \pm 0.47$, which is derived from near-UV observations of a sample of Galactic and extra-Galactic novae in outburst \citep{44} (see Fig. \ref{fig:2}). Moreover, a Li yield per nova event of $A(Li) = 7.1$ is about what is estimated to make the Li abundance presently observed \citep{45}. Finally, the amount of Lithium measured from optical spectroscopy of V1369 Cen, $M_{Li} = (2.6 \pm 2.2) \times 10^{-10}$ M$_{\odot}$ \citep{18}, corresponds to only 8.7$\%$ of the total, considering the epoch of the spectrum (namely, $t = 7$ days) from the nova explosion, and the half-life decay time of $^7$Be, T$_{1/2} = 53.12 \pm 0.06$ days. Consequently, based on optical spectroscopy performed at the epoch of the nova outburst, the amount of total $^7$Be synthesized during the TNR in V1369 Cen going to enrich the interstellar medium is $M_{^7Be} = (3.0 \pm 2.5) \times 10^{-9}$ M$_{\odot}$ (see also Appendix). This is equivalent to a yield of $N(Li)_{opt} = 6.5 \pm 0.4$, which is consistent within 1 $\sigma$ with the value obtained from the analysis of the 478 keV line (Fig. \ref{fig:2}).

\section{Conclusions}

In this work, we present possible evidence of the $^7$Be line at 478 keV, as predicted by \citet{7}, and arising from the decay of beryllium-7 to lithium via electron capture. Despite extensive searches, this line has remained undetected until now \citep{8}. The emission was observed by the INTEGRAL satellite during the explosion of V1369 Cen, the brightest nova observed so far this century. 

The possible detected emission exhibits a flux of $F = (4.9 \pm 2.0) \times 10^{-4}$ ph/cm$^2$/s, which corresponds to a 2.5$\sigma$ confidence level. Although indicative of potential gamma-ray activity at 478 keV, this significance level remains below the threshold required to assert an unequivocal detection. The flux excess is centered at 479.0 $\pm$ 2.5keV and is temporally and spatially coincident with the outburst of V1369 Cen. At a distance of $d = 643(+405, -112)$ pc, determined using multiple methods, including observations from the \textit{Gaia} satellite \citep{33}, this flux corresponds to a total $^7$Be mass of $M = (1.2^{+2.0}_{-0.6}) \times 10^{-8}$ M$_{\odot}$. This value is higher than the average $^7$Be mass typically produced in nova events and is sufficient to account for the full amount of lithium estimated by \citet{45}. By incorporating the total ejected mass of V1369 Cen, as determined from radio observations \citep{40}, the atomic fraction of $^7$Be=Li in the outburst is calculated to be $A(Li) = 7.1^{+0.7}_{-0.3}$. 

The analysis of the abundance obtained from the 478 keV line from $^7$Be decay aligns with previous $^7$Be and $^7$Li results obtained with near-UV and optical spectroscopy using ground-based telescopes in all novae where Li has been searched, see Fig. \ref{fig:2}, solidifying novae as main Li producers in the Milky Way. However, the derived Li abundances exceed theoretical predictions by a full order of magnitude, further highlighting the discrepancy with TNR calculations \citep{46,47,48}.

%%%%%%%%%%%%%%%%%%%%%%%%%%%%%%%%%%%%%%%%%%%%%%%%%%%%%%%%%%%%%%
\begin{acknowledgements}
      We want to thank the anonymous referee for their valuable comments and suggestions that greatly contributed to improving the quality of this manuscript. We also greatly appreciate Margarita Hernanz and Carme Jordi for precious discussions and clarifications that have improved the structure of the manuscript. We also warmly thank Brad Schaefer for important discussions related to the Gaia distance to V1369 Cen and Jurgen Knodlseder for the support in the analysis of INTEGRAL/SPI data. The INTEGRAL/SPI project has been completed under the responsibility and leadership of CNES; we are grateful to ASI, CEA, CNES, DLR, ESA, INTA, NASA and OSTC for support of this ESA space science mission. PB acknowledge support from the ERC advanced grant N. 835087 – SPIAKID. LI acknowledges financial support from the YES Data Grant Program (PI: Izzo) {\it Multi-wavelength and multi messenger analysis of relativistic supernovae}. We acknowledge with thanks the variable star observations from the AAVSO International Database contributed by observers worldwide and used in this research. {\it Data availability.} The INTEGRAL data used in this manuscript are publicly available on the INTEGRAL Cosmos website, hosted by the European Space Agency (ESA): https://www.cosmos.esa.int/web/integral/ integral-data-archives. The optical spectra of V1369 Cen are publicly available on the European Southern Observatory science archive facility: https://archive.eso.org/cms.html. The Python notebooks used in this analysis will be available in a dedicated repository hosted on GitHub publicly-available personal page of the first author: https://github.com/lucagrb/V1369Cen

\end{acknowledgements}

% Please note that we have included the references below in
% order to compile the document, but we ask you to:

%%%%%%%%%%%%%%%%%%%%%%%%%%%%%%%%%%%%%%%%%%%%%%%%%%%%%%%%%%%%%%%
% Appendices must be placed after   \end{thebibliography}
% They will be placed automatically on a new page.
%%%%%%%%%%%%%%%%%%%%%%%%%%%%%%%%%%%%%%%%%%%%%%%%%%%%%%%%%%%%%%%
\begin{appendix}
%%%%%%%%%%%%%%%%%%%%%%%%%%%%%%%%%%%%%%%%%%%%%%%%%%%%%%%%%%%%%%%
% In the PDF output, floats should be placed
% under their own appendix, not before the title, nor after the
% title of the next appendix.

% In short appendices, onecolumn floats (\figure*
% or \table*) will generate a blank page.
% To prevent this behaviour, a few examples are provided here. 

% In case you have a lot of floating objects for little text and the 
% LaTeX engine moves the floats away from their context, the command
% \FloatBarrier of the “placeins” package will empty the
% float buffer and place all stored floats in the continuity.

% If you still encounter problems with wide floats placement,
% just use the onecolumn environment throughout the appendices.
%%%%%%%%%%%%%%%%%%%%%%%%%%%%%%%%%%%%%%%%%%%%%%%%%%%%%%%%%%%%%%%

\section{An accurate estimate of the lithium mass measured from the Li I 670.8 nm \ line}

Here we revisit the measurement of lithium abundance in V1369 Cen \citep{18}, using the curve of growth methodology, widely used to estimate physical properties, especially abundances, of an absorbing medium along the light of sight between the observer and the emitting source \citep{50}, which in this case if the pseudo-photosphere from the underlying V1369 nova outburst. Thanks to the high resolution provided by FEROS, we have clearly identified and resolved the transitions from Li I, as well as from Na I and K I, namely elements that share with lithium the same electronic configuration in their most external orbitals (all of them are alkali metals), similar excitation energies for their ground state transitions and that they have been observed in their dominant ionization state. However, the structure of V1369 Cen ejecta does not allow for a single fit for the entire absorption lines using a Gaussian or a Voigt model; this is particularly true for the Na ID doublet. In these cases, it is customary to use the equivalent width $W$ of the entire absorption line, which is indeed a net measurement of the fraction of energy removed from the spectral continuum by the absorbing element in the ejecta, and then by the absorption line under consideration. From the equation of the radiative transport, and the assumption that the ejecta can be modeled as a thermal plasma with a given Maxwellian distribution described by a Doppler parameter $b$, we have that the specific equivalent width $W_{\lambda}$ is proportional to \citep{50}:

\begin{equation}
    W_{\lambda} = \frac{W}{\lambda} = \frac{2 b}{c} F\left(\frac{1.497 \times 10^{-2}}{b} N_i f_{ij} \lambda\right),
\end{equation}
where $\lambda$ is the wavelength of the line transition under consideration, $f_{ij}$ is the oscillator strength of the transition, $N_i$ is the column density. The function $F$ can be numerically integrated providing a relation between the specific equivalent width and the column density for a given Doppler parameter $b$: the curve of growth. 

Consequently, from an accurate measurement of the equivalent widths for the above-mentioned ground state transitions, it is possible to derive simultaneously the Doppler parameter $b$ and the corresponding column densities. However, large column densities imply large optical depths, and then partial or complete saturation of the absorption line. When the absorption line is affected by saturation, the curve of growth starts to flatten, with the main consequence that a small variation in the equivalent width implies a large variation, and uncertainty, in the resulting column density, once $b$ is determined. 

We have used the Day 7 epoch spectrum to measure equivalent widths for ground state transitions of Na I D1 and D2 lines, the K I 769.9 nm, and Li I 670.8 nm lines. The K I  lines are located in a region heavily affected by telluric lines. We have then performed a telluric correction by computing the telluric correction directly from the science spectrum, using a line-by-line radiative transfer model (LBLRTM\footnote{\url{http://rtweb.aer.com/lblrtm.html}}) with atmospheric input extracted from the science spectrum file header. This code attempts to best fit the observed spectrum by varying the composition of the atmosphere (water vapor and $O_2$ ), as well as the pressure and the temperature to take into account possible variations within the total exposure time. However, despite this treatment, the K I 766.5 nm line profile cannot be fully recovered, given the presence of heavily saturated telluric lines at the same wavelengths of the P-Cygni absorption originating in the nova ejecta. Equivalent widths have been obtained by performing the following measurement for each absorption line:
\begin{equation}
    \Sigma_{i} \left(1 - \frac{I_{c,i}}{I_i}\right)
\end{equation}
where $i$ represents the single pixel wavelength (measured  in \AA), with $I_{c,i}$ the continuum flux and $I_i$ the observed flux at the pixel $i$. 

\begin{figure}
    \centering
    \includegraphics[width=0.7\linewidth]{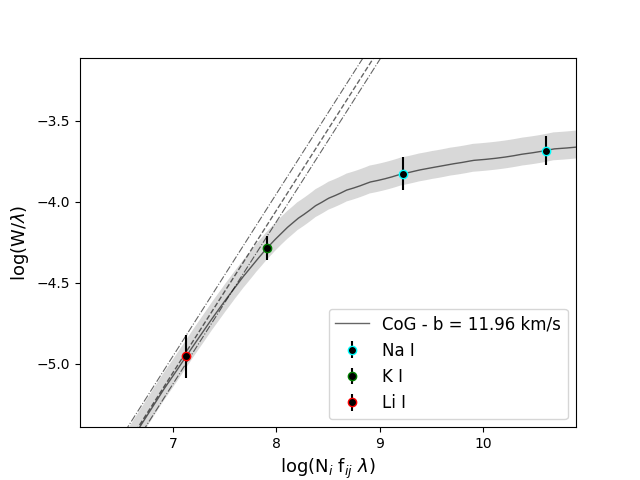}
    \includegraphics[width=0.7\linewidth]{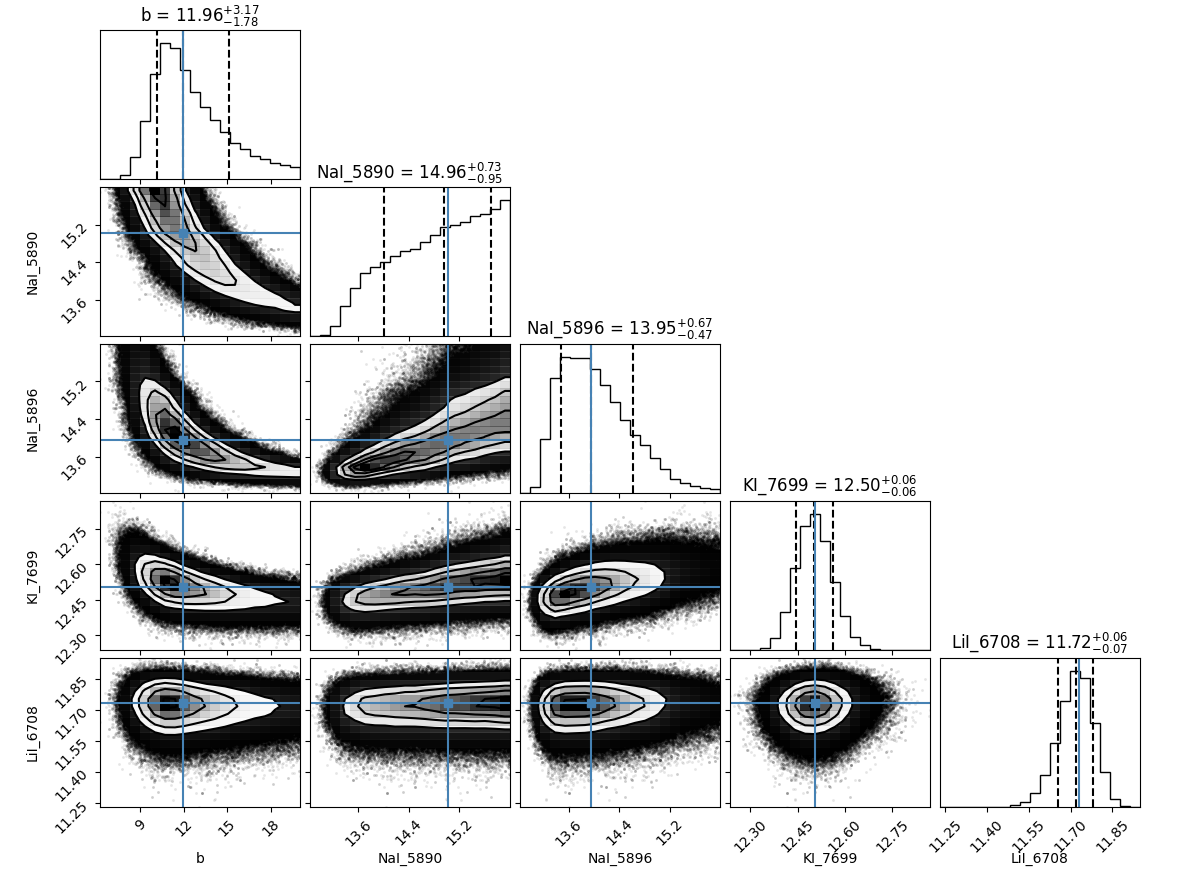}
    
    \caption{(Upper panel) 
    The best-fit curve of growth obtained from the equivalent width values measured in the Day 7 spectrum of V1369 Cen. Na I D lines are located in the flat region of the curve of growth suggesting that they are saturated and hardly usable for precise abundance estimates (see also Fig. \ref{fig:CoG}). On the other hand, Ki and Li I lines are located in the linear region. this allowed us to estimate the lithium over-abundance and infer the amount of lithium mass in the nova ejecta from the Li I 670.8 nm absorption line. (Lower panel) The posterior distribution for the Doppler parameter $b$ and the column densities $N_i$ derived from the MCMC procedure applied to the curve of growth best fit.}
    \label{fig:CoG}
\end{figure}

To estimate the column density of lithium, we have developed a procedure that first performs a simultaneous best fit to search for the Doppler parameter $b$ and the column density values of sodium, potassium, and lithium using their detected ground state transitions. The latter transition is a doublet, but separating the two lines is also difficult for a high-resolution spectrograph like FEROS, so we here have considered the Li I 670.8 nm feature as a single line. Then, we performed a Monte Carlo Markov Chain analysis, using the {\em emcee} ensemble sampler python package\footnote{\url{https://emcee.readthedocs.io/en/stable/}}, to estimate the posterior distributions, and then uncertainties, of the above parameters, obtaining the results displayed in Fig. \ref{fig:CoG}. The curve of growth corresponding to the best-fit $b = 11.99$ km/s, with the column densities derived for each single transition using this methodology, is shown in Fig. \ref{fig:CoG}. An immediate conclusion from this analysis is that sodium lines are saturated, located on the flat region of the curve of growth, and their best-fit column densities are not very precise. On the other hand, lithium $\left( \log(N_{Li,cen}/cm^{-2}) = 11.76^{+0.10}_{-0.13}\right)$ and potassium $\left( \log(N_{K,cen}/cm^{-2}) = 12.28^{+0.08}_{-0.09}\right)$ column densities are very well precise, being located in the linear region of the curve of growth, fig. \ref{fig:CoG}. Using only potassium as the reference element, we get an abundance ratio $N_{Li}/N_K = 30/100$, and after correcting for the differential ionization of lithium of 0.54 \citep{51,52} and adopting a solar abundance of $N(KI) = 5.12 \pm 0.07$ \citep{53} we obtain an absolute  $N(Li) = 5.14 \pm 0.10$. 

Finally, we must consider that the total amount of $^7$Be that has already decayed into lithium on Day 7, namely when our abundance estimate has been performed, is provided by 
\begin{equation}
    M_{^7Be} = 10^{N(Li) - 12} \times u_{Li} \times M_{ej} / (2^{\frac{7}{T_{1/2}}} - 1) 
\end{equation}

where $T_{1/2}$ is the half-life time decay of $^7$Be, and $u_{Li} = 7$. This value is 8.7$\%$ of the total initial abundance of beryllium, implying that the initial total abundance of lithium in the ejecta of V1369 Cen, as measured from the Li I 670.8 nm line, is $N(Li) = 6.5 \pm 0.4$. This value is in agreement with the respective uncertainties with the estimate obtained through the detection of the $^7$Be 478 keV line.

%____________________________________________________________
%       Wide floats at the start of an appendix: first method
%-------------------------------------------------------------
% To prevent a blank page after the start of an appendix:
% - Switch to one \onecolumn first
% - Declare the section title
% - Declare the onecolumn float with the parameter [h!]
% - Revert to \twocolumn at the end of the section
\onecolumn

\section{{\em Gaia} stars used for the estimate of V1369 distance using the DIB 862.1 nm}

\begin{table}[h!]
    \centering
\begin{tabular}{l|c|c|c|c}
\hline\hline
         {\em Gaia} DR3 ID &   RA (J2000.0) &   Dec (J2000.0) & EW               & Distance            \\
\hline
 5870610893673250688 &        208.789 &        -59.1888 & 0.122 $\pm$ 0.041 & 1538.84 $\pm$ 29.83  \\
 5870611026755720448 &        208.832 &        -59.1626 & 0.1 $\pm$ 0.017   & 1857.12 $\pm$ 65.53  \\
 5870604262243142912 &        208.544 &        -59.2671 & 0.075 $\pm$ 0.02  & 2435.92 $\pm$ 84.74  \\
 5870657107526303488 &        208.458 &        -59.0804 & 0.16 $\pm$ 0.091  & 1935.52 $\pm$ 66.1   \\
 5870662948682465792 &        208.656 &        -59.0033 & 0.051 $\pm$ 0.011 & 1119.48 $\pm$ 19.28  \\
 5870603368890103552 &        208.704 &        -59.2998 & 0.073 $\pm$ 0.015 & 372.36 $\pm$ 2.32    \\
 5870612474221731328 &        209.025 &        -59.0833 & 0.034 $\pm$ 0.118 & 947.71 $\pm$ 13.43   \\
 5870605705294226432 &        208.371 &        -59.3016 & 0.099 $\pm$ 0.068 & 905.46 $\pm$ 11.48   \\
 5870623812936215680 &        209.136 &        -59.125  & 0.079 $\pm$ 0.013 & 1400.06 $\pm$ 31.8   \\
 5870595053832590976 &        209.077 &        -59.3115 & 0.186 $\pm$ 0.147 & 1366.25 $\pm$ 54.68  \\
 5870589590633520768 &        208.681 &        -59.4575 & 0.092 $\pm$ 0.127 & 2382.27 $\pm$ 133.98 \\
 5870669648831799168 &        208.759 &        -58.804  & 0.042 $\pm$ 0.006 & 681.07 $\pm$ 8.36    \\
 5870638106588762752 &        209.277 &        -58.9711 & 0.041 $\pm$ 0.055 & 790.27 $\pm$ 8.89    \\
 5870510872466795776 &        208.215 &        -59.4116 & 0.369 $\pm$ 0.009 & 3439.06 $\pm$ 382.39 \\
 5870621339034647168 &        209.436 &        -59.1433 & 0.219 $\pm$ 0.059 & 2367.3 $\pm$ 120.65  \\
 5870494860827168640 &        208.369 &        -59.4997 & 0.082 $\pm$ 0.053 & 1078.84 $\pm$ 16.9   \\
 5870564645462040960 &        207.908 &        -59.1826 & 0.253 $\pm$ 0.037 & 3507.5 $\pm$ 257.18  \\
 5870683044762906112 &        208.572 &        -58.7429 & 0.042 $\pm$ 0.005 & 702.6 $\pm$ 6.43     \\
 5870581202499273984 &        207.855 &        -59.0472 & 0.16 $\pm$ 0.014  & 3662.9 $\pm$ 244.44  \\
 5870506268261657600 &        208.177 &        -59.5221 & 0.088 $\pm$ 0.045 & 1089.61 $\pm$ 22.79  \\
 5867580742686508928 &        208.68  &        -59.6451 & 0.054 $\pm$ 0.027 & 256.84 $\pm$ 1.25    \\
\hline\hline
\end{tabular}
\caption{The list of {\em Gaia} DR3 stars surrounding V1369 Cen used for the determination of the distance through DIB 862.1 nm EW (column 4) and their {\em Gaia} distance (column 5).}
    \label{tab:GaiaGS}
\end{table}

% it is easier to finish the page in onecolumn and revert to
% twocolumn when starting the next page if needed.}

\end{appendix}
\end{document}